\documentclass{article}

\RequirePackage[numbers,sort&compress]{natbib}
\usepackage{float}
\usepackage{hyperref}
\hypersetup{
    colorlinks=true,
    urlcolor=cyan
}

\RequirePackage{graphicx,xcolor}
\RequirePackage{colortbl}
\RequirePackage{booktabs}
\RequirePackage{algorithm,amssymb,amsmath}
\RequirePackage[noend]{algpseudocode}
\RequirePackage{changepage}
\RequirePackage[twoside,letterpaper,includeheadfoot,layoutsize={8.125in,10.875in},layouthoffset=0.1875in,layoutvoffset=0.0625in,left=38.5pt,right=43pt,top=43pt,bottom=32pt,headheight=0pt,headsep=10pt,footskip=25pt]{geometry}

\usepackage[T1]{fontenc}
\usepackage[utf8]{inputenc}
\usepackage{authblk}

\usepackage[font={small}]{caption}

\usepackage[printwatermark]{xwatermark}
\usepackage{tikz}

\title{The role of industry, occupation, and location specific knowledge in the survival of new firms}

\author[a]{C. Jara-Figueroa}
\author[a]{Bogang Jun}
\author[b]{Edward Glaeser}
\author[a,1]{Cesar Hidalgo}
\affil[a]{MIT Media Lab, Massachusetts Institute of Technology}
\affil[b]{Department of Economics, Harvard University}
\affil[1]{To whom correspondence should be addressed. E-mail: hidalgo@mit.edu}

\begin{document}\sloppy

\twocolumn[
  \begin{@twocolumnfalse}
\maketitle
\begin{abstract}
How do regions acquire the knowledge they need to diversify their economic activities? How does the migration of workers among firms and industries contribute to the diffusion of that knowledge? Here we measure the industry, occupation, and location specific knowledge carried by workers from one establishment to the next using a dataset summarizing the individual work history for an entire country. We study pioneer firms--firms operating in an industry that was not present in a region--because the success of pioneers is the basic unit of regional economic diversification. We find that the growth and survival of pioneers increase significantly when their first hires are workers with experience in a related industry, and with work experience in the same location, but not with past experience in a related occupation. We compare these results with new firms that are not pioneers and find that industry specific knowledge is significantly more important for pioneer than non-pioneer firms. To address endogeneity we use Bartik instruments, which leverage national fluctuations in the demand for an activity as shocks for local labor supply. The instrumental variable estimates support the finding that industry related knowledge is a predictor of the survival and growth of pioneer firms. These findings expand our understanding of the micro-mechanisms underlying regional economic diversification events.
\end{abstract}
  \end{@twocolumnfalse}]

Can developing countries and cities thrive through their own entrepreneurship, or must they attract external investment? What are the factors that influence the success of local ventures? Development depends on undertaking new tasks, which require knowledge. In this paper, we estimate the impact of a worker's knowledge about an industry, occupation, and location in the survival of pioneer firms \cite{hausmann2016workforce}: firms that start operating in a region where their industry was not present.

Understanding the success of pioneer firms is key to understanding the mechanisms behind industrial diversification. When a pioneer firms succeeds, the region where this firm is now present will have successfully developed a new industry. Here, we use a large administrative data set with almost complete work histories for all the individual workers of a country, to measure the knowledge carried by workers from their previous jobs into pioneer firms. This dataset allows us to estimate the industry specific knowledge, occupation specific knowledge, formal schooling, and knowledge about a location that each worker brings into a pioneer firm. We use this fine grained description to to test which type of knowledge matters most for the growth and survival of pioneer firms, and compare these results with new firms that are not pioneers; non-pioneer firms. 

For decades, human capital has been recognized as an important determinant of economic growth \cite{Romer1990,Nelson1966investment,Barro1991,glaeser1992growth,Rauch1993,Glaeser1994cities,Glaeser2000new, Barro2001humancapital, Gennaioli2012human}. But human capital is not just a worker's formal schooling. Workers acquire important skills, knowledge, and contacts at work. A forty-year-old worker brings, on average, more years of experience into a company than years of schooling. This work experience, which is specific to an industry, location, and occupation, should impact the growth and survival of the activities where these workers are involved. 

The specificity of this knowledge pushes us to think of human capital not only in terms of intensity, but in terms of relatedness. Workers are not simply knowledgeable or skilled, but posses knowledge that is related to specific activities, even to new activities that have never before been present in a city or a country. In this paper, we test what type of related knowledge is a more critical ingredient in the success of new firms that lead to the development of new industries. While there is a long literature measuring relatedness between products \cite{hidalgo2007product, hausmann2014atlas}, industries \cite{Neffke2011, NeffkeHenning2013}, technologies \cite{Kogler2013, boschma7geography}, and even occupations \cite{muneepeerakul2013urban}, there is little work separating these relatedness measures into multiple forms of human capital.

\begin{figure*}[h!t]
    \centering
    \includegraphics[width=.95\linewidth]{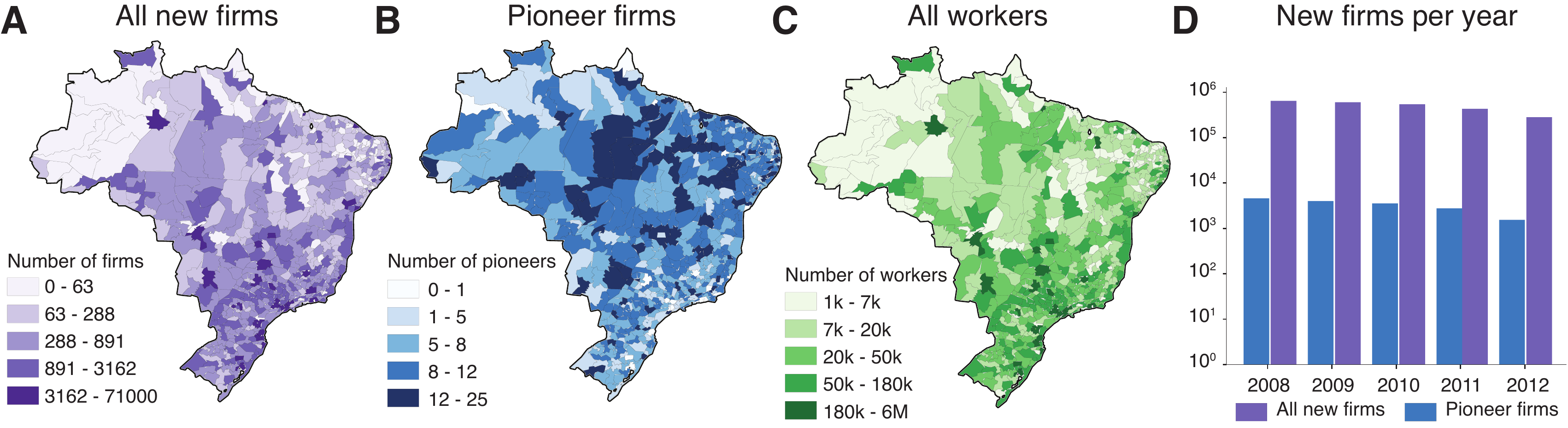}
	\caption{Spatial distribution of new firms in Brazil created between 2008 and 2012. \textbf{A} all firms, \textbf{B} only pioneer firms, and \textbf{C} distribution of workers. \textbf{D} number of firms created each year. } \label{fig:map}
	\vspace{-10pt}
\end{figure*}

In this paper we decompose knowledge into a two-dimensional representation, measuring how related the previous experience of a worker is to the industry and to the occupation of their new job. A worker with abundant formal schooling and experience can be classified as someone with little related experience if her work history involves occupations and industries that are unrelated to her current employment. Conversely, a worker with low formal education can be classified as having high related experience if she moves into an industry and occupation that are related to the ones she has performed previously. The dimensions of industry and occupation knowledge are not necessarily tied together, since a worker can have abundant experience in the occupation of her new job, while having very little experience in a related industry. We test the relative importance of these dimensions of knowledge relatedness for the survival and growth of pioneer firms, and compare these results with their relative importance for new firms that are not pioneers. 

The idea that workers bring in knowledge into the firms they participate in is an idea that has a long tradition in organizational learning. According to Herbert Simon, organizations acquire knowledge either by the learning of its members or by ingesting new members \cite{Simon1991}. Because pioneer firms do not start with members that can learn, the knowledge this firm has needs to come from the workers that it hires. We find that the survival of pioneer firms increases significantly when their first hires are people with industry specific knowledge, and with experience in that location, but not with occupation specific knowledge. When comparing pioneers with non-pioneers, we find that industry knowledge is significantly more important for pioneers than for non-pioneers, and that occupation specific knowledge plays a relatively more important role for non-pioneers.

There are some serious concerns relating to the endogeneity of starting a firm and of hiring. For instance, firms with more social capital may be able to hire more people from related industries. We cannot address these concerns fully, but we can instrument for the number of workers from a related industry available in a labor market by looking at national industrial shifts using a Bartik-style instrument \cite{bartik1991}. Intuitively, the supply of related workers is higher in areas with related local industries that have received adverse national or global shocks. Our results on the importance of related knowledge are similar when we use this instrument.

Together, our results show how work histories can be used to measure the types of knowledge brought by workers into pioneer firms, and also, help uncover the relative importance of industry and occupation specific knowledge in pioneering economic activities. These results tell us that the success of the pioneering activities that promote diversification depends strongly on the move of local workers with related knowledge into these new activities.

\section{Data}

We use Brazil's RAIS (Annual Social Security Information Report) compiled by the Ministry of Labor and Employment (MET) of Brazil between 2002 and 2013. The RAIS dataset uses the National Classification of Economic Activities (CNAE) for industries, and the Brazilian Occupations Classification (CBO) for occupations, both revised by the Brazilian Institute of Geography and Statistics (IBGE).

The RAIS dataset covers about 97\% of the Brazilian formal labor market \cite{cardoso2007international} and contains fine-grained information about individual workers, including 5,570 municipalities (which are grouped by the IBGE into 558 microregions based on similar productive structure and spatial interaction \cite{IBGE1990}), 501 occupations, and 284 industries for more than 30 million workers each year. Location information is provided at the discrete level of each municipality, so a continuous treatment is not possible. Municipalities in Brazil are grouped by IBGE into microregions based on similar productive structure and spatial interaction \cite{IBGE1990}. Microregions are grouped into 137 mesoregions, which are grouped into 27 states, and states are grouped into 5 macroregions. All the results presented in the main text use the 3-digit level for industries, the 4-digit level for occupations, and microregions as the spatial unit of analysis. We use microregions because they provide a more stringent criteria than municipalities for identifying pioneer firms; it is easier to be the first firm to operate in an industry inside a small municipality than inside a much larger microregion. The supplementary information provides an alternative operational definition of pioneer firms based on microregions plus their neighborhood. 


One of the key characteristics of RAIS that make it so useful for research is its granularity. The variables in RAIS can be tracked down to the individual level, which makes it the most important source of information on the formal labor market dynamics in the country. The classification of industries went through a major revision between 2005 and 2006, which we solve by splitting the analysis into before and after 2006.

\begin{figure*}[ht]
    \centering
    \includegraphics[width=.95\linewidth]{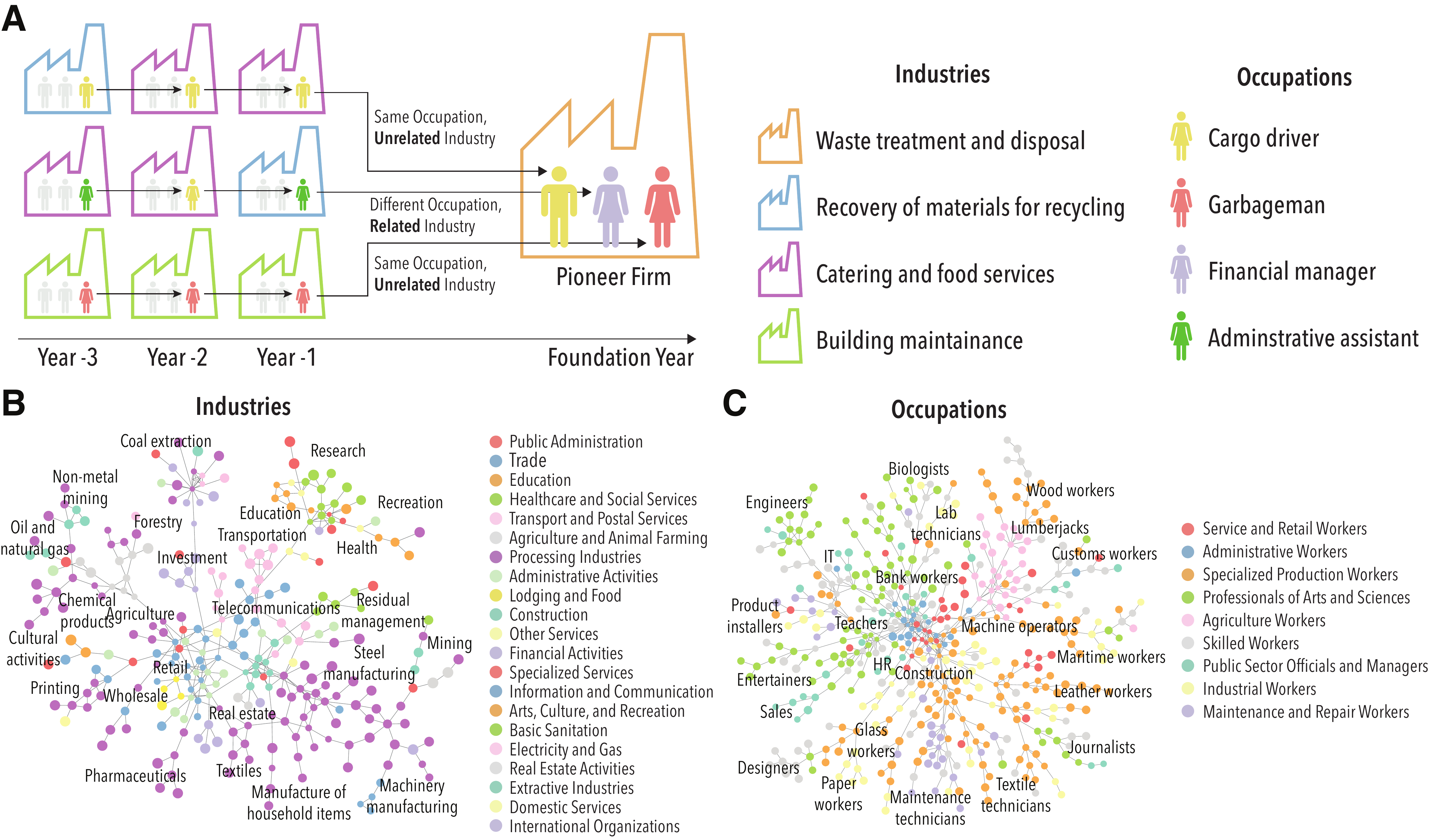}
	\caption{Work histories and networks of related activities. The diagram in \textbf{A} shows how individual work histories are used to infer the knowledge brought into the pioneer firm by its first hires. The color of each worker represents their occupation, while the color of the bounding box represents the industry. The yellow worker, for example, has experience as a cargo driver, the same occupation he was hired to perform in the pioneer firm, but comes from a very unrelated industry. The light blue worker has experience in a different occupation, but in a related industry. \textbf{B} shows the network of related industries and \textbf{C} shows the network of related occupations \textbf{C}. Node colors correspond to the highest level of the classification for occupations and industries. This figure only shows the most important edges for each network, selected based on a trimming algorithm that starts with the maximum spanning tree and then adds all edges above a threshold (see \href{https://dam-prod.media.mit.edu/uuid/c53345af-1dd8-4216-902d-36b75b126f83}{SI Appendix} for details). } \label{fig:diagram}
	\vspace{-10pt}
\end{figure*}

Unfortunately, a firm that does not declare RAIS in a particular year may not be necessarily ``dead,'' but just facing economic problems that make it rational not to pay taxes in that year or not to appear in any official control mechanism. In fact, many firms simply freeze their activities awaiting better economic events. This will lead us to underestimate the survival rate of firms, although the exit from RAIS is surely itself an important event. Because Brazilian legislation makes it relatively easy to open a company, but relatively difficult to close one, many firms, especially small firms, often close without informing official authorities, suggesting that the exit from RAIS might be a better expression of a company's status than the official closing of the firm. Studies conducted by the IBGE and MTE estimate that the rate of underreport of firms' death range from 14\% to 20\% of actually closed firms. To partially address these issues, we will consider firms to be ``dead'' when they stop reporting for at least two consecutive years. Despite these limitations, RAIS is the main source of information on the rate of firm creation and destruction at the municipal level \cite{cardoso2007international}. In fact, the Central Registry of Firms (CEMPRE) is built by IBGE and MTE based on the information available in RAIS. 

\section{Results}

Pioneer firms are the basic units of economic diversification. Here, we define a pioneer firm as a firm that is new (no record of it for at least 6 years), and that operates in an industry that is new to its region (no record of the industry in the region for at least two years before the pioneer). For companies starting after 2006 we will add the extra condition that they operate for at least two consecutive years, so as to filter out small short lived firms. Because we need at least two years of work history of the a pioneer's first hires, and because CNAE went through a major revision between 2005 and 2006, we analyze only firms created either in 2005, or after 2008 (for more information see \href{https://dam-prod.media.mit.edu/uuid/c53345af-1dd8-4216-902d-36b75b126f83}{SI Appendix}). 

Figure \ref{fig:map} shows the spatial distribution for all new firms (A), pioneer firms (B), and workers (C), across Brazilian microregions between 2008 and 2012. During the observation period, Brazil produced roughly 500,000 new firms a year, of which only about 3,000 to 4,000 (less than 1\%) were pioneers (Figure \ref{fig:map} D). For information about the industries of pioneer firms see \href{https://dam-prod.media.mit.edu/uuid/c53345af-1dd8-4216-902d-36b75b126f83}{SI Appendix}. 

For pioneers, all their employees are new hires, so all their initial stock of knowledge is connected to their initial workforce \cite{Simon1991}. We base our measure of the knowledge brought in by a company's new hire on the industry and the occupation of their previous job. Because of the limited time range of the data, we consider only jobs performed during the two years before the creation of the pioneer firm. For instance, if a worker was a teller (occupation) for a telecommunication company (industry), we assume that she brings two types of knowledge to the pioneer firm: industry specific knowledge about the telecommunication industry and occupation specific knowledge about being a teller. Because different industries and different occupations vary along a continuum, we abandon the view of industry and occupation knowledge as two binary variables \cite{castanias2001managerial}. We instead use a continuous approach building on the literature on relatedness. For example, the industries of shoe manufacturing and shirt manufacturing are different industries, but they are similar enough that a worker moving from shoe manufacturing to shirt manufacturing should be regarded as having some industry specific knowledge about shirt manufacturing, relative to workers coming from a less related industry such as animal agriculture. The diagram presented in Figure \ref{fig:diagram} A shows a pioneer firm made of three workers: the first and third come from the same occupation, but an unrelated industry, and the second comes from a different occupation, but a related industry. 

\begin{figure*}[ht]
    \centering
    \includegraphics[width=.95\linewidth]{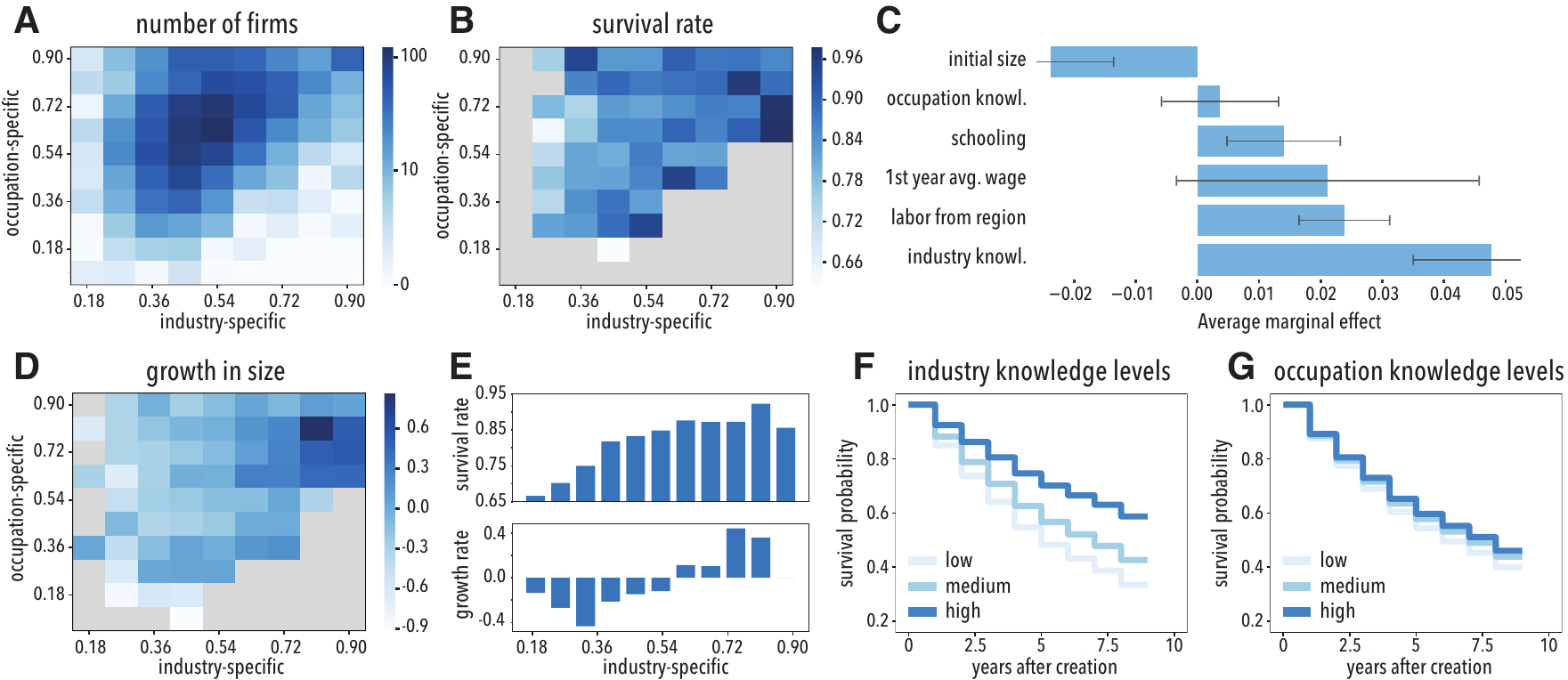}
	\caption{Characteristics of pioneer firms that started after 2008, as a function of the industry and occupation specific knowledge brought by their workers: \textbf{A} shows the number of firms observed in the data, \textbf{B} shows the empirical survival rate at the third year, \textbf{D} shows the empirical employment growth rate at the third year of firms that survived, and \textbf{E} shows survival rate and growth rate as a function of industry specific knowledge only. The gray color represents situations with not enough data points. \textbf{C} Shows the average Marginal Effect on survival of each variable from model (6) in Table \ref{tab:regModelSurvivalGrowth}, for firms that started after 2008. \textbf{F} Shows the predicted values for model (5) from Table \ref{tab:Cox2005} for firms that started in 2005, for different levels of industry knowledge: low, medium, and high. \textbf{G} is similar to F, but for different levels of occupation knowledge. In both F and G, \textit{low} means the smallest observed value among pioneers, \textit{medium} means the median of the observed values, and \textit{high} means the maximum observed value.} \label{fig:heatmap}
	\vspace{-10pt}
\end{figure*}

To measure the relatedness between the industry of a pioneer firm and the work histories of that firm’s workers we follow the literature on relatedness and use labor flows between pairs of industries at the national level \cite{Neffke2011,NeffkeHenning2013}. Similarly, we measure relatedness for each pair of occupations by looking at labor flows among occupations across the entire Brazilian economy. Unfortunately, the CBO classification has not been successfully linked to skill compositions, so we cannot use direct measure of skill similarity. Logically, labor should flow freely between industries and occupations that require similar knowledge and not between industries and occupations that require wildly different knowledge. In fact, the relatedness measure based on labor mobility has been termed ``skill relatedness'' by some authors \cite{NeffkeHenning2013,delgado2015defining}, because individuals changing jobs will likely remain in activities that value the skills associated with their previous work.

Formally, we define the relatedness between industry $i$ and industry $i'$ as the residual of a regression explaining labor flows as a function of the size of industries and their growth rates \cite{NeffkeHenning2013}. That is, we consider a pair of industries (occupations) to be related when the labor flows between them is higher than what we would expect based on the size and growth of a pair of industries. In other words, we take the residuals of the regression from Eq. \ref{eq:indFlowReg}, where $F_{i \leftrightarrow i'}^{(t)}$ is the total flow of workers in log-scale going from $i$ to $i'$ and from $i'$ to $i$ between year $t-1$ and $t$. $g_{ii'}^{(t)} = \max \{ g_i^{(t)} , g_{i'}^{(t)} \}$ is the maximum growth rate in the number of employees $g_i^{(t)} = \ln L_i^{(t)} - \ln L_i^{(t-1)}$ between both industries, $\tilde{L}^{(t)}_{ii'} = \max \{ L_i^{(t)} , L_{i'}^{(t)} \}$ is the maximum number of employees between both industries, in log-scale, and $L_i^{(t)}$ is the number of employees of industry $i$ in year $t$, also in log-scale. We normalize the residuals $\hat{\gamma}_{ii'}^{(t)}$ to keep them between zero and one (see Eq. \ref{eq:indResNorm}). We measure relatedness between occupations $o$ and $o'$ in an analogous way (see Eqs. \ref{eq:occFlowReg} and \ref{eq:occResNorm}). 

\begin{align}
    F_{i \leftrightarrow i'}^{(t)} = & \  \beta_0 + \beta_1 g_{ii'}^{(t)} + \beta_2 \tilde{L}^{(t)}_{ii'} + \gamma_{ii'}^{(t)},\label{eq:indFlowReg}\\
    \phi_{ii'}^{(t)} = & \  \left\lbrace \begin{matrix}
    \frac{\hat{\gamma}_{ii'}^{(t)} - \min_{ii'} \{ \hat{\gamma}_{ii'}^{(t)} \}}{\max_{ii'} \{ \hat{\gamma}_{ii'}^{(t)} \}-\min_{ii'} \{ \hat{\gamma}_{ii'}^{(t)} \} } &,& i\ne i'\\
    1 &,& i=i'
    \end{matrix}
    \right.\label{eq:indResNorm}
\end{align}

\begin{align}
    F_{o \leftrightarrow o'}^{(t)} = & \  \beta_0 + \beta_1 g_{oo'}^{(t)} + \beta_2 \tilde{L}^{(t)}_{oo'} + \theta_{oo'}^{(t)},\label{eq:occFlowReg}\\
    \psi_{oo'}^{(t)} = &\  \left\lbrace \begin{matrix}
    \frac{\hat{\theta}_{oo'}^{(t)} - \min_{oo'} \{ \hat{\theta}_{oo'}^{(t)} \}}{\max_{oo'} \{ \hat{\theta}_{oo'}^{(t)} \}-\min_{oo'} \{ \hat{\theta}_{oo'}^{(t)} \} } &,& o \ne o'\\
    1 &,& o=o'
    \end{matrix}
    \right.\label{eq:occResNorm}
\end{align}

Relatedness among industries and among occupations define two weighted undirected networks for each year. Figures \ref{fig:diagram} B and C show the networks of related industries and occupations for 2008, after selecting the most important edges for purpose of the visualization (see \href{https://dam-prod.media.mit.edu/uuid/c53345af-1dd8-4216-902d-36b75b126f83}{SI Appendix} for details). All of our analysis are conducted with the full, time dependent, weighted networks. 

Next, we use these measures of relatedness to create indicators of the stock of related knowledge that workers bring into pioneer firms. For each pioneer firm, we measure the amount of industry and occupation specific knowledge brought into it by its workers by aggregating relatedness across all its workers:
\begin{align}
    \Phi_{f,i,r}^{(t)}=& \ \sum_{i'} s_{f,i'} \phi_{ii'}^{(t)}\\ \Psi_{f,i,r}^{(t)}=& \ \sum_{o'} s_{f,o,o'} \psi_{oo'}^{(t)},
\end{align}
where $s_{f,i'}$ is the fraction of workers in firm $f$ with experience on industry $i'$, and $s_{f,o,o'}$ is the fraction of workers in firm $f$ performing occupation $o$ with experience in occupation $o'$.

These two aggregate variables quantify, respectively, the industry and occupation specific knowledge that workers bring--based on their previous experience--into a pioneer firm $f$. 

\begin{table*}[!ht] \centering 
\tiny
\setlength{\tabcolsep}{1pt}
\begin{tabular}{lcccccccccccc} 
 & \multicolumn{12}{c}{\textit{Dependent variable:}} \\ 
\cline{2-13} 
\\[-1.8ex] & \multicolumn{6}{c}{Survival rate at third year, $S^{(t+3)}$} & \multicolumn{6}{c}{Three year growth rate, $G^{(t+3)}$ } \\ 
\\[-1.8ex] & (1) & (2) & (3) & (4) & (5) & (6) & (7) & (8) & (9) & (10) & (11) & (12)\\ 
\hline \\[-1.8ex] 
 Industry knowl. &  & 0.466$^{***}$ &  &  &  & 0.457$^{***}$ &  & 0.174$^{***}$ &  &  &  & 0.185$^{***}$ \\ 
 \ \ \ \ \   ($\Phi$) &  & (0.114) &  &  &  & (0.123) &  & (0.029) &  &  &  & (0.031) \\ 
 Occupation knowl. &  &  & 0.184$^{**}$ &  &  & 0.035 &  &  & 0.033 &  &  & $-$0.029 \\ 
 \ \ \ \ \  ($\Psi$) &  &  & (0.085) &  &  & (0.092) &  &  & (0.022) &  &  & (0.023) \\ 
 Years of schooling &  &  &  & 0.163$^{*}$ &  & 0.134 &  &  &  & 0.023 &  & 0.012 \\ 
 \ \ \ \ \ ($edu$)  &  &  &  & (0.086) &  & (0.091) &  &  &  & (0.025) &  & (0.025) \\
 Local knowledge &  &  &  &  & 0.238$^{***}$ & 0.228$^{***}$ &  &  &  &  & 0.014 & 0.007 \\ 
 \ \ \ \ \ ($\rho$) &  &  &  &  & (0.071) & (0.072) &  &  &  &  & (0.019) & (0.019) \\ 
 Initial size & $-$0.246$^{***}$ & $-$0.251$^{***}$ & $-$0.261$^{***}$ & $-$0.226$^{**}$ & $-$0.235$^{**}$ & $-$0.227$^{**}$ & $-$0.393$^{***}$ & $-$0.394$^{***}$ & $-$0.395$^{***}$ & $-$0.391$^{***}$ & $-$0.393$^{***}$ & $-$0.391$^{***}$ \\ 
\ \ \ \ \  ($\log(n_0)$)  & (0.093) & (0.095) & (0.094) & (0.092) & (0.093) & (0.096) & (0.031) & (0.030) & (0.031) & (0.031) & (0.031) & (0.030) \\ 
 Average wage & 0.208 & 0.136 & 0.188 & 0.137 & 0.342 & 0.202 & 0.231$^{***}$ & 0.209$^{***}$ & 0.228$^{***}$ & 0.221$^{***}$ & 0.238$^{***}$ & 0.208$^{***}$ \\ 
  \ \ \ \ \ ($\log (w)$)  & (0.220) & (0.233) & (0.221) & (0.224) & (0.235) & (0.257) & (0.071) & (0.069) & (0.071) & (0.072) & (0.072) & (0.071) \\ 
\hline \\[-1.8ex] 
Year f.e. & \checkmark & \checkmark & \checkmark & \checkmark & \checkmark & \checkmark & \checkmark & \checkmark & \checkmark & \checkmark & \checkmark & \checkmark \\ 
Industry f.e. & \checkmark & \checkmark & \checkmark & \checkmark & \checkmark & \checkmark & \checkmark & \checkmark & \checkmark & \checkmark & \checkmark & \checkmark \\ 
Region f.e. & \checkmark & \checkmark & \checkmark & \checkmark & \checkmark & \checkmark & \checkmark & \checkmark & \checkmark & \checkmark & \checkmark & \checkmark \\ 
\hline \\[-1.8ex] 
Observations & 1,632 & 1,632 & 1,632 & 1,632 & 1,632 & 1,632 & 1,376 & 1,376 & 1,376 & 1,376 & 1,376 & 1,376 \\ 
McFadden & 0.2128 & 0.2265 & 0.2161 & 0.2153 & 0.2212 & 0.2367 &  &  &  &  &  &  \\ 
AICc & 1,635.9 & 1,619.1 & 1,633.9 & 1,635.0 & 1,626.6 & 1,612.7 &  &  &  &  &  &  \\ 
Log Likelihood & $-$558.1 & $-$548.4 & $-$555.8 & $-$556.3 & $-$552.1 & $-$541.1 &  &  &  &  &  &  \\ 
R$^{2}$ &  &  &  &  &  &  & 0.324 & 0.343 & 0.325 & 0.324 & 0.324 & 0.344 \\ 
Adjusted R$^{2}$ &  &  &  &  &  &  & 0.194 & 0.216 & 0.194 & 0.194 & 0.194 & 0.215 \\
F Statistic &  &  &  &  &  &  & 2.490$^{***}$ & 2.699$^{***}$ & 2.487$^{***}$ & 2.481$^{***}$ & 2.480$^{***}$ & 2.665$^{***}$ \\ 
&  &  &  &  &  &  & (df = 222) & (df = 223) & (df = 223) & (df = 223) & (df = 223) & (df = 226) \\ 
\hline \\[-1.8ex] 
\textit{Note:}  & \multicolumn{12}{r}{$^{*}$p$<$0.1; $^{**}$p$<$0.05; $^{***}$p$<$0.01 and standard errors are in parentheses.}
\end{tabular} 
\caption{\textmd{Estimates of the effect of different types of knowledge on the survival rate (models 1-6, logistic regressions) and growth rate (models 7-12, OLS) at the third year for pioneer firms. For all models reported standard errors are robust and clustered by region, and the four knowledge variables are expressed in standard deviation units.}} \label{tab:regModelSurvivalGrowth} 
\vspace{-15pt}
\end{table*} 

Figure \ref{fig:heatmap} A shows a bi-variate histogram of the number of pioneer firms starting with a certain stock of industry and occupation specific knowledge. We note that the median relatedness between a pair of industries or a pair of occupations is about 0.4, so most pioneer firms hire workers with a level of industry and occupation relatedness that is much higher than if they would be hiring those workers at random. The best interpretation of this fact is that the firms and workers recognize the importance of related knowledge and search and hire accordingly. When we study the histogram we observe that pioneer firms tend to hire workers with occupation specific knowledge (top rows) but only with an intermediate level of industry specific knowledge (middle columns).

Next, we look at the pioneer firms that survive. Figure \ref{fig:heatmap} B shows a bi-variate histogram for the average three year survival rate of pioneer firms. Surprisingly, the distribution of surviving firms is quite different from the distribution of all pioneer firms. While pioneer firms tend to hire workers with occupation specific knowledge, surviving pioneer firms tend to be those that hired workers with high levels of industry specific knowledge (Figure \ref{fig:heatmap} B). In fact, the three year survival rate of pioneer firms increases from about 60\% when workers do not have industry specific knowledge, to more than 85\% when workers bring an average industrial relatedness of more than $\Phi_{f} > 0.5$ (Figure \ref{fig:heatmap} E). Figure \ref{fig:heatmap} D shows the growth in employment of surviving pioneer firms. Here we see that pioneer firms with high stocks of industry specific knowledge also grow much faster than those lacking industry specific knowledge (Figure \ref{fig:heatmap} E). 

We formalize these results using multivariate regression analysis that predicts the three year survival rate $S_{f,i,r}^{(t+3)}$ and employment growth $G_{f,i,r}^{(t+3)}$ of pioneer firm $f$, operating in industry $i$ and region $r$. We use logistic regression to predict the three year survival rate and OLS to predict growth. We focus on the three year survival rate as a simple way to address right censoring of our data (companies that outlive our observation period). If we were to study survival at longer time periods using a logistic model, we would have to shrink the pool of pioneer firms we can track (for alternative models see \href{https://dam-prod.media.mit.edu/uuid/c53345af-1dd8-4216-902d-36b75b126f83}{SI Appendix}).

Our models for survival and growth are a function of the firm's stock of industry specific knowledge ($\Phi$), occupation specific knowledge ($\Psi$), average years of schooling of its workers ($edu$), number of initial workers ($n_{0}$), average wage ($w$), and local knowledge ($\rho$), which we define as the fraction of workers with work experience in the same region. In all of our models, the four knowledge variables ($\Phi,\Psi,edu,\rho$) are measured in units of standard deviations from their respective means, to make their coefficients more easily interpretable and comparable. Formally, our models take the form defined in Eqs. \ref{eq:survivalModel} and \ref{eq:growthModel}. The model in Eq. \ref{eq:survivalModel} is a logistic regression, and $\mu_i$, $\lambda^{(t)}$, and $\eta_r$ from Eqs. \ref{eq:survivalModel} and \ref{eq:growthModel} are industry, year, and region fixed effect, respectively. Because we control for these fixed effects, our model can capture the effect of different types of human capital on firms' survival and growth, while controlling for time-invariant characteristics of industries and regions (such as the life cycle of an industry), as well as nation wide trends. Moreover, by adding the initial number of workers and the average wage of each firm, we are controlling for size effects and for the other effects regarding how attractive the jobs at each firm are. 

\begin{table*}[!ht] \centering 
  \tiny
\setlength{\tabcolsep}{5pt}
\begin{tabular}{lcccccc} 
 & \multicolumn{6}{c}{\textit{Dependent variable:}} \\ 
\cline{2-7} 
\\[-1.8ex] & \multicolumn{3}{c}{Survival rate at third year, $S^{(t+3)}$} & \multicolumn{3}{c}{Three year growth rate, $G^{(t+3)}$} \\ 
\\[-1.8ex] & (1) & (2) & (3) & (4) & (5) & (6)\\ 
\hline \\[-1.8ex] 
Industry knowl. & 0.457$^{***}$ & 0.091$^{***}$ & 0.091$^{***}$ & 0.185$^{***}$ & 0.054$^{***}$ & 0.054$^{***}$ \\ 
\ \ \ \ \ ($\Phi$) & (0.123) & (0.007) & (0.007) & (0.031) & (0.002) & (0.002) \\
Pioneer dummy &  &  & 0.156 &  &  & 0.088$^{**}$ \\ 
  &  &  & (0.126) &  &  & (0.038) \\ 
Industry knowl.:pioneer dummy  &  &  & 0.203$^{**}$ &  &  & 0.091$^{***}$ \\ 
  &  &  & (0.093) &  &  & (0.027) \\ 
Occupation knowl. & 0.035 & 0.035$^{***}$ & 0.036$^{***}$ & $-$0.029 & 0.012$^{***}$ & 0.011$^{***}$ \\ 
\ \ \ \ \ ($\Psi$)   & (0.092) & (0.007) & (0.007) & (0.023) & (0.002) & (0.002) \\ 
Years of schooling & 0.134 & 0.008 & 0.009 & 0.012 & 0.002 & 0.002 \\ 
 \ \ \ \ \ ($edu$) & (0.091) & (0.007) & (0.007) & (0.025) & (0.002) & (0.002) \\ 
Local knowledge & 0.228$^{***}$ & 0.084$^{***}$ & 0.085$^{***}$ & 0.007 & $-$0.007$^{***}$ & $-$0.007$^{***}$ \\ 
\ \ \ \ \  ($\rho$) & (0.072) & (0.006) & (0.006) & (0.019) & (0.002) & (0.002) \\ 
\hline \\[-1.8ex] 
Firm controls & \checkmark & \checkmark & \checkmark & \checkmark & \checkmark & \checkmark \\ 

Year f.e. & \checkmark & \checkmark & \checkmark & \checkmark & \checkmark & \checkmark \\ 
Industry f.e. & \checkmark & \checkmark & \checkmark & \checkmark & \checkmark & \checkmark \\ 
Region f.e. & \checkmark & \checkmark & \checkmark & \checkmark & \checkmark & \checkmark \\ 
Firm type & pioneers & non-pioneers & all & pioneers & non-pioneers & all \\
\hline \\[-1.8ex] 
Observations & 1,632 & 284,369 & 286,001 & 1,376 & 242,192 & 243,568 \\ 
McFadden & 0.2367 & 0.0404 & 0.0404 &  &  &  \\ 
AICc & 1,613 & 231,739 & 233,106 &  &  &  \\ 
Log Likelihood & $-$541 & $-$115,638 & $-$116,320 &  &  &  \\ 
R$^{2}$ &  &  &  & 0.344 & 0.152 & 0.152 \\ 
Adjusted R$^{2}$&  &  &  & 0.215 & 0.151 & 0.151 \\ 
F Statistic &  &  &  & 2.665$^{***}$  & 188.475$^{***}$  & 188.274$^{***}$  \\ 
&  &  &  & (df = 226) & (df = 230) & (df = 232) \\
\hline \\[-1.8ex] 
\textit{Note:}  & \multicolumn{6}{r}{$^{*}$p$<$0.1; $^{**}$p$<$0.05; $^{***}$p$<$0.01 and standard errors are in parentheses.} \\ 
\end{tabular} 
\caption{\textmd{Survival and growth at the third year for pioneer firms (models 1 and 4), non-pioneer firms (models 2 and 5), and for all new firms (models 3 and 6). The interaction between industry knowledge and a dummy for pioneers is positive and significant, meaning that the effect of industry specific knowledge is larger for pioneer companies. As before, all knowledge variables are expressed in standard deviation units. Firm controls include initial size and average wage.}} \label{tab:regModelSurvivalGrowthAll}
\vspace{-15pt}
\end{table*}

\begin{align}
    S^{(t+3)}_{f,i,r} = &\  \beta_0 +  \beta_1 \Phi_{f,i,r}^{(t)} + \beta_2 \Psi_{f,i,r}^{(t)} + \beta_3 edu_{f,i,r}^{(t)} + \beta_4 \rho_{f,i,r}^{(t)}  \nonumber \\
    & + \beta_5 \log(n_{0\ f,i,r}^{(t)}) + \beta_6 \log(w_{f,i,r}^{(t)}) \nonumber \\
    & + \mu_i + \lambda^{(t)} + \eta_r + \varepsilon_{f,i,r}^{(t)}  \label{eq:survivalModel}\\
    G_{f,i,r}^{(t+3)} = &\  \beta_0 +  \beta_1 \Phi_{f,i, r}^{(t)} + \beta_2 \Psi_{f,i,r}^{(t)} + \beta_3 edu_{f,i,r}^{(t)} + \beta_4 \rho_{f,i,r}^{(t)} \nonumber \\
&  + \beta_5 \log(n_{0\  f,i,r}^{(t)}) + \beta_6 \log( w_{f,i,r}^{(t)}) \nonumber \\
& + \mu_i + \lambda^{(t)} + \eta_r + \varepsilon_{f,i,r}^{(t)}  \label{eq:growthModel}
\end{align}

Table \ref{tab:regModelSurvivalGrowth} presents the results for both models for pioneer firms, with $\Phi$, $\Psi$, $edu$, and $\rho$ measured in standard deviation units. Across all specifications the effects of industry specific knowledge ($\Phi$) in the survival and growth of firms remains strong, whereas the effects of occupation specific knowledge ($\Psi$) and schooling ($edu$), are weak when considered in isolation, and insignificant after controlling for industry specific knowledge ($\Phi$). Figure \ref{fig:heatmap} C shows the average marginal effects for model (6) from Table \ref{tab:regModelSurvivalGrowth}. An increase in one unit of standard deviation of industry knowledge leads to an average $\sim 5\%$ increase in the firm's probability of survival.

Is industry knowledge important only for pioneer firms, or for all new firms? Table \ref{tab:regModelSurvivalGrowthAll} shows a comparison between pioneers and other non-pioneer new firms. The industry knowledge coefficient for non-pioneers is significantly lower than for pioneers (the interaction term in model (3) is positive and significant), and for non-pioneers the occupation knowledge coefficient remains significant even when we consider it together with industry specific knowledge. Although we cannot reject the view that general knowledge and occupation related knowledge matter for both pioneers and for all firms, our results show that their effect is small compared to industry specific knowledge. In fact, the point estimate for schooling is actually larger for pioneers than for all new firms. These results suggest that industry specific knowledge is more important for pioneer firms than for new firms.

To explore the long-run impact of knowledge on survival, we focus on firms that started operating in 2005 and use the Cox Proportional Ratios model \cite{bender2005generating, cox2018analysis} with a similar specification as before (Eq. \ref{eq:survivalModel}). Since we are only using pioneers from one year, a fixed effects model would lead to model overspecification. Instead, we control for region and firm characteristics as shown in Table \ref{tab:Cox2005}. Figures \ref{fig:heatmap} F and G show the predicted values for the survival rate of pioneer firms according to model (5) from Table \ref{tab:Cox2005}, for firms with low, medium, and high level of industry knowledge (Figure \ref{fig:heatmap} F) and occupation knowledge (Figure \ref{fig:heatmap} G). Industry knowledge has more distinctive effects on the survival rate than occupation specific knowledge (more details in \href{https://dam-prod.media.mit.edu/uuid/c53345af-1dd8-4216-902d-36b75b126f83}{SI Appendix}).

\begin{table}[!htbp] \centering 
  \tiny
\setlength{\tabcolsep}{1pt}
\begin{tabular}{lccccc} 
 & \multicolumn{5}{c}{\textit{Dependent variable:}} \\ 
\cline{2-6} 
\\[-1.8ex] & \multicolumn{5}{c}{death probability} \\ 
\\[-1.8ex] & (1) & (2) & (3) & (4) & (5) \\ 
\hline \\[-1.8ex] 
Industry knowl. & $-$0.214$^{**}$ &  &  &  & $-$0.181$^{**}$ \\ 
\ \ \ \ \ ($\Phi$) & (0.089) &  &  &  & (0.092)  \\ 
  Occupation knowl. &  & $-$0.107$^{*}$ &  &  & $-$0.038  \\ 
 \ \ \ \ \ ($\Psi$) &  & (0.059) &  &  & (0.063)  \\
  Years of schooling  &  &  & $-$0.129$^{**}$ &  & $-$0.105$^{*}$  \\ 
  \ \ \ \ \ ($edu$) &  &  & (0.057) &  & (0.058) \\ 
  local knowledge  &  &  &  & $-$0.145$^{***}$ & $-$0.144$^{***}$  \\ 
 \ \ \ \ \ ($\rho$)  &  &  &  & (0.047) & (0.048) \\ 
\hline \\[-1.8ex] 
 Region controls  & \checkmark & \checkmark & \checkmark & \checkmark & \checkmark  \\ 
 Firm controls & \checkmark & \checkmark & \checkmark & \checkmark & \checkmark  \\ 
\hline \\[-1.8ex] 
Observations  & 462 & 462 & 462 & 462 & 462 \\ 
R$^{2}$ & 0.026 & 0.019 & 0.023 & 0.032 & 0.054  \\ 

Wald Test  & 11.580  & 9.070& 10.790 & 15.660$^{**}$ & 25.840$^{***}$  \\ 
  & (df = 8) &  (df = 8) & (df = 8) &  (df = 8) & (df = 11)  \\ 

\hline \\[-1.8ex] 
\textit{Note:}  & \multicolumn{5}{r}{$^{*}$p$<$0.1; $^{**}$p$<$0.05; $^{***}$p$<$0.01} \\ 
\end{tabular} 
\caption{\textmd{Cox proportional hazards model for pioneer firms that started on 2005. Firm controls include initial size and average wage, and region controls include population, GDP per capita, average schooling, available industry knowledge, and the survival rate of non-pioneer firms as a control for how competitive the region is. As before, all knowledge variables are expressed in standard deviation units.}} \label{tab:Cox2005} 
\vspace{-15pt}
\end{table}

The endogeneity of firm entry and hiring decisions both challenge these results. Perhaps, more productive firms just tend to hire related industry workers. Perhaps, occupation related knowledge does not matter, because firms only enter when they anticipate their ability to make up for any lack in occupation related skill. We cannot address all endogeneity concerns, but we use shocks to the supply of related human capital at the local level as an instrument of hiring such workers. 

Here, we construct a Bartik labor supply shock $B_{ri}$ \citep{diamond2016determinants, blanchard1992regional, bartik1991} using the demand shocks experienced by other related industries. In other words, we use the growth or decline of industry $i'$ at the national level, as a supply shock that respectively decreases or increases the availability of the workers with industry specific knowledge required by industries related to $i'$. For instance, if the manufacturing of cars and motorcycles are related in terms of industry specific knowledge, a demand boom in the car sector would cause a shortage of workers with knowledge relevant to the manufacturing of motorcycles in the regions where the car industries are growing. Consequently, we should expect a pioneer firm in the motorcycle industry to hire less workers with industry specific knowledge when the industries related to motorcycle manufacturing are experiencing national level booms. This means the expected correlation, through this mechanism, between the Bartik instrument $B_{ri}$ and the number of workers with industry specific knowledge hired by a pioneer firm $\Phi_{f}$ should be \textit{negative}.

We define the industry knowledge Bartik shock on industry $i$ in region $r$ as:
\begin{align}
B_{ri}^{(ind)}(t) = &\  \sum_{i', i'\ne i} g_{i';r}^{(t)} \frac{\phi_{ii'}^{(t)} L_{ri'}^{(t)}}{\sum_{i', i'\ne i} \phi^{(t)}_{ii'}L_{ri'}^{(t)}}\label{EQ:BartikSupplyInd},
\end{align} 
where $\phi_{ii'}^{(t)}$ is the relatedness between industries $i$ and $i'$, using flows between $t-1$ and $t$, $g_{i;r}^{(t)} = \log(L_{i;r}^{(t)}) - \log(L_{i;r}^{(t-1)})$ is the employment growth of industry $i$ in every region except in region $r$, and $L_{i;r}^{(t)}$ is the number of workers in year $t$ in industry $i$ removing region $r$. $L_{ri'}^{(t)}$ is the number of people working on industry $i'$ in region $r$. Eq. \ref{EQ:BartikSupplyInd} has the same form as the original Bartik shock, since it is an interaction between the national trend ($g_{i';r}^{(t)}$) with the local industrial structure ($L_{ri'}^{(t)}$), but weighted by the similarity with industry $i$ ($\phi_{ii'}^{(t)}$).

Table \ref{tab:Bartik} shows the results of using $B_{ri}^{(ind)}$ as an instrument for industry knowledge $\Phi$ to estimate the effect of industry knowledge in the growth of pioneer firms. Our two-stage least squares estimates confirm the sign of the effect found using OLS. 

\begin{table}[!htbp] \centering 
  \tiny
\setlength{\tabcolsep}{1pt}
\begin{tabular}{lcccc} 
 & \multicolumn{4}{c}{\textit{Dependent variable:}} \\ 
\cline{2-5} 
\\[-1.8ex] & Industry knowl.& \multicolumn{3}{c}{Three year growth rate} \\ 
\\[-1.8ex] & \textit{First} & \textit{Reduced} & \textit{Instrumental} & \textit{OLS} \\ 
 & \textit{stage} & \textit{form} & \textit{variable} & \textit{} \\ 
\\[-1.8ex] & (1) & (2) & (3) & (4)\\ 
\hline \\[-1.8ex] 
 Industry knowl. &  &  & 0.502$^{**}$ & 0.177$^{***}$ \\ 
 \ \ \ \ \ ($\Phi^{(t)}$) &  &  & (0.256) & (0.032) \\ 
 Bartik shock & $-$6.899$^{***}$ & $-$3.465$^{**}$ &  &  \\ 
  
\ \ \ \ \ ($B_{ri}^{(ind)}$) & (1.568) & (1.686) &  &  \\ 
 Growth of industry  & 0.282$^{**}$ & $-$0.003 & $-$0.144 & 0.056 \\ 
 \ \ \ \ \ ($g_{i;r}$) & (0.134) & (0.144) & (0.162) & (0.161) \\ 
 Constant & $-$0.634$^{***}$ & 0.496$^{***}$ & 0.814$^{***}$ & 1.898$^{***}$ \\ 
  & (0.075) & (0.081) & (0.243) & (0.549) \\ 
\hline \\[-1.8ex] 
Observations & 1,380 & 1,380 & 1,380 & 1,380 \\ 
R$^{2}$ & 0.016 & 0.003 &   & 0.234 \\ 
Adj. R$^{2}$ & 0.015 & 0.002 &  & 0.089 \\  
F Statistic & 11.236$^{***}$  & 2.129  &  & 1.609$^{***}$ \\ 
& (df = 2) & (df = 2) &  &  (df = 220) \\ 
\hline \\[-1.8ex] 
\textit{Note:}  & \multicolumn{4}{r}{$^{*}$p$<$0.1; $^{**}$p$<$0.05; $^{***}$p$<$0.01 } \\ 
\end{tabular} 
\caption{\textmd{Results of using the Bartik shock defined in Eq. \ref{EQ:BartikSupplyInd} as an instrument for the industry specific knowledge brought to a pioneer firm by its first hires ($\Phi$). Our two stage least squares estimates confirm the direction of the effect on growth found using OLS. The F-test for the strength of the instrument yields a statistic of 18.339$^{***}$ \cite{stock2002testing}. Industry knowledge is expressed in standard deviation units. }} \label{tab:Bartik} 
\vspace{-5pt}
\end{table}

\section{Discussion}

Here we use the entire work history of Brazil to create measures for the knowledge carried by workers into new activities and study how these different types of knowledge affect the growth and survival of pioneer firms. Pioneer firms--new firms operating in an industry that is new for the region--are of particular interest because their success represents an increase in regional economic diversification. Our work shows that industry specific knowledge is particularly important, since pioneer firms that hire workers with experience in a related industry grow faster and are more likely to survive. Surprisingly, the effect of occupation specific knowledge and general schooling are not significant for pioneer firms, while being important for newly formed non-pioneer firms.

Knowledge diffusion is acknowledged to be a key driver of economic development. In fact, countries and cities have been shown to be more likely to develop new economic activities that are similar to their existing activities \cite{hidalgo2007product,boschma2017relatedness,Neffke2011,NeffkeHenning2013,BoschmaFrenken2011}. This effect has proven so strong that, at the international level, less than 8 percent of the recorded diversification events between 1970 and 2010 were into unrelated products \cite{pinheiro2018}. Yet, most research on industrial diversification has focused on the macro-level dynamics. Here we contribute to this body of literature by studying the micro-level mechanisms that might lead to this type of observations \cite{dopfer2004micro}.

The idea that workers carry the knowledge that economies need to grow and diversify is not new. Yet, knowledge and human capital are usually conceptualized as measures of intensity (years of schooling for example). Our evidence suggest that knowledge is better understood in terms of relatedness since workers differ not only in their total knowledge, but also in what this knowledge is about. Here we have shown that general knowledge, measured as average years of schooling, is not a strong determinant of the survival of a pioneer firm, but that the relatedness of knowledge between past and present activities is.

Moreover, we show that for pioneer firms, industry knowledge is a stronger predictor of survival and growth than occupational knowledge. This is an unexpected finding. One explanation for this might be that the first hires of a pioneer company often end up taking some managerial role, while not operating directly as managers. For these roles, industry specific knowledge might be more important than occupation specific knowledge. Another possible explanation could be simply that industry-specific skills take longer to acquire than occupation-specific skills, and hence, firms with more in-house industry experience have an advantage at the outset. 

Imagine the case of a salesperson. Salespeople are essential for the growth and survival of firms and have both occupation and industry specific knowledge. The occupation specific knowledge of a salesperson involves knowledge on how to communicate with clients, develop relationships, and close deals. These are skills that can be easily transferred from one firm to the next. The industry specific knowledge required by a salesperson, however, depends strongly on the product or service being sold. A salesperson with experience in selling garments may struggle selling enterprise software, not because she cannot develop a relationship with a client, but because she may lack the knowledge needed to understand the software needs of clients and the engineering capacity of her team. Lacking the experience needed to understand and communicate needs precisely, a salesperson without industry specific knowledge can generate misunderstandings between clients and production teams that could be disastrous for a pioneer company. 

Previous work has shown that the founder's experience is a strong predictor of the performance of start-ups \cite{shane2003}. We do not know who the founder of the company is in our data, but we can check whether the observed effect is due to just one employee or if it is a characteristic of the team. We find that an important part of the effect is driven by the most experienced (related) employee, but that there is a significant part that is due to the rest of the team. Even after we remove the most experienced member of the team from the sample and add her as a pioneer specific control, our finding that industry specific knowledge matters remains strong. This suggests that the most experienced employee is not driving all of the observed effect (see \href{https://dam-prod.media.mit.edu/uuid/c53345af-1dd8-4216-902d-36b75b126f83}{SI Appendix}). 

Another explanation for our results is that workers from related industries are more likely to have connections to clients, customers, and trustworthy workers, so what they bring is not just their knowledge about the industry, but also their knowledge of the social network where the industry is embedded in \cite{granovetter1985economic,uzzi1997social}. This form of industry specific social capital, can be regarded as a sub-type of industry specific knowledge or experience, and also, should be reflected in the locations specific knowledge of a worker, which we find is a significant predictor of the growth and survival of pioneer firms. Unfortunately, there are few data sources that can be used to isolate the effects of skills and location with the pure effects of social capital, so the effects of embeddedness are hard to identify. 

These findings add to the literature studying differences between industry and occupation specific knowledge in other contexts \cite{Neal1995industry,Gibbons2004task}. The industry knowledge brought by a firm's manager, for example, has been shown to be very important for the productivity of the firm \cite{castanias2001managerial,bailey2003external}. In fact, a manager's human capital has been shown to be mainly industry specific \cite{Sullivan2010}, in the sense that industry tenure provides a higher wage premium than occupational tenure. For other occupations such as craftsmen, human capital has been shown to be primarily occupation specific. Together with this body of literature, our study suggests that the picture where a job (an occupation for a given industry) is linked to a set of skills only through the occupation might be incomplete. 

There is growing evidence of the effects of movement of industry specific human capital on the development of regions. History shows that the migration of skilled workers encourages regional development of new industries. For example, in the sixteenth century, the region around Antwerp was an industrial center for the textile industry, until the anti-Protestant persecution in the late sixteenth century triggered an exodus of Protestant workers. Many of those skilled workers moved to the northern part of the Netherlands and helped develop new textile industries in those cities \cite{israel1990empires,obrien2001urban}. Similarly, other studies using pioneer plants have revealed the importance of industry specific human capital \cite{hausmann2016workforce}, but have not compared it with general knowledge or occupational knowledge. 

Although our data is specific to Brazil, the great variation in income and industrialization level among Brazilian microregions suggests that our results might generalize. In fact, the richest Brazilian microregion had an average income per capita in 2013 of about USD 28k, which was comparable to that of Spain, Italy, or South Korea; while the poorest microregions had an average income of about USD 5k, which is comparable to that of Paraguay, Jamaica, or Algeria. Moreover, the vast geographic variation of wealth in Brazil makes it an interesting scenario for studying industrial development, since it combines the challenges of middle income countries with the data reporting quality of high income countries. 
Finally, our results emphasize that in order to fully understand the importance of tacit knowledge for regional industrial diversification it is important to measure knowledge along different dimensions. The work history of individuals may be the key to measure these different types of knowledge.

\section*{Acknowledgements}

\small{
C.J.F., B.J., and C.A.H. acknowledge support from the MIT Media Lab Consortia, the MIT Skoltech Program, the Masdar Institute of Science and Technology (Masdar Institute), Abu Dhabi, USA—Reference 02\slash MI\slash MIT\slash CP\slash 11\slash 07633\slash GEN\slash G\slash 00 and the Center for Complex Engineering Systems (CCES) at King Abdulaziz City for Science and Technology (KACST). We thank comments from Kerstin Enflo, Jian Gao, Tarik Roukny, and Siqi Zheng, as well as data assistance from Manuel Aristar\'an and Elton Freitas.
}

\bibliography{reference}

\end{document}